\documentclass[aps,superscriptaddress,eqsecnum,nofootinbib,preprintnumbers]{revtex4}
\usepackage{graphicx,epsfig}
\usepackage{amsmath}
\usepackage{amssymb}
\usepackage{subfigure}

\usepackage{relsize}

\newcommand{\be}{\begin{equation}}
\newcommand{\ee}{\end{equation}}
\newcommand{\bea}{\begin{eqnarray}}
\newcommand{\eea}{\end{eqnarray}}

\begin{document}

\title{Is the use of Christoffel connection in gravity theories conceptually correct?
}

\author{Georgios Kofinas}
\email{gkofinas@aegean.gr} \affiliation{Research Group of Geometry,
Dynamical Systems and Cosmology,\\
Department of Information and Communication Systems Engineering,\\
University of the Aegean, Karlovassi 83200, Samos, Greece}

\begin{abstract}

Christoffel connection did not enter gravity as an axiom of minimal length for the free
fall of particles (where anyway length action is not defined for massless particles), nor out of
economy, but from the weak equivalence principle (gravitational force is equivalent to acceleration
according to Einstein) together with the identification of the local inertial frame with the
local Lorentz one. This identification implies that the orbits of all particles are given by the
geodesics of the Christoffel connection. Here, we show that in the presence of only massless
particles (absence of massive particles) the above identification is inconsistent and does
not lead to any connection. The proof is based on the existence of projectively equivalent
connections and the absence of proper time for null particles. If a connection derived by some
kinematical principles for the particles is to be applied in the world, it is better these principles
to be valid in all relevant spacetime rather than different principles to give different connections
in different spacetime regions. Therefore, our result stated above may imply a conceptual
insufficiency of the use of the Christoffel connection in the early universe where only massless
particles are expected to be present (whenever at least some notions, like orbits, are meaningful),
and thus of the total use of this connection. If in the early universe the notion of a massive
particle, which appears latter in time, cannot be used, in an analogous way in a causally
disconnected high-energy region (maybe deep interior of astrophysical objects or black holes) the
same conclusions could be extracted if only massless particles are present.

\end{abstract}

\maketitle


\section{Introduction}

Christoffel connection is undoubtedly the dominant connection in gravitational physics.
For positive-definite metrics of Riemannian geometry the geodesics of the Christoffel connection provide
the curves of minimal (sometimes maximal) length as is manifested by the variation of the length action
$\int d\sigma\sqrt{g_{\mu\nu}\frac{dx^{\mu}}{d\sigma}\frac{dx^{\nu}}{d\sigma}}$ for some parameter
$\sigma$ along the orbits $x^{\mu}(\sigma)$. Similarly, in spacetime physics, the action
$\int d\sigma\sqrt{-g_{\mu\nu}\frac{dx^{\mu}}{d\sigma}\frac{dx^{\nu}}{d\sigma}}$ extremizes the proper
time (or the length in some sense) of timelike curves, which are the curves of
massive particles, and provides the geodesics of the Christoffel connection as the
corresponding orbits. However, for massless particles the proper time is zero, a photon orbit
on the null cone has zero length, so a similar action is not defined.
Therefore, the axiom of minimum length for massive particles is accomplished by another axiom,
that of continuation of the geodesic equation of massive particles to be also valid in the massless
case. This fact does not mean that the geodesic equation for all particles does not arise
from an action. It arises e.g. from the action
$\int d\sigma \big(eg_{\mu\nu}\frac{dx^{\mu}}{d\sigma}\frac{dx^{\nu}}{d\sigma}-e^{-1}m^{2}\big)$,
where $m$ is the particle mass and $e$ a one-dimensional vierbein. This last action is meaningful
even in the massless case, where now the nullity constraint
$g_{\mu\nu}\frac{dx^{\mu}}{d\sigma}\frac{dx^{\nu}}{d\sigma}=0$
provides through multiplication with the Lagrange multiplier $e$ a meaningful (off-shell) Lagrangian
to be varied (actually in agreement with the method of the theory of constraints). However, this last
action is still an axiom with purpose to get the geodesic equation of the Christoffel connection
as providing the orbits of all particles. In a similar way the minimal surface action
of string theory is assumed. The fact that the Christoffel connection is
the simplest one, which only contains the metric part and not extra degrees
due to torsion or non-metricity, does not necessarily declare it as the physical connection to
describe the universe (anyway, our job is not topography to stick to the notion of minimal length).
A lot of modifications of General Relativity contain various extra fields in the action without
enriching the Christoffel connection with extra fields, therefore the economy argument of the
Christoffel connection is immediately canceled.
Moreover, people today are more concerned about the particle content of gravity theories
built up from the Christoffel or other connections and rarely follow Einstein's spirit that principles
on particle kinematics could provide a connection to built up a theory. In a sense, string theory
is closer to this spirit since it starts with the motion of the string to construct the theory.
At the opposite end is the first order formulation of various gravity theories, where the metric (or
the vierbein) and the connection are considered quite independent; the cost however from this
approach is that the connection (or equivalently the torsion or non-metricity) carry in general
a large number of new components and it is difficult to believe that all of them have some physical
meaning. An intermediate situation probably looks more promising.

Christoffel connection entered physics through Einstein, who certainly did not adopt it
axiomatically or in a geometric manner, but he had to be convinced for its virtue out of
physical arguments. Of course, one can still keep on using the Christoffel connection
even without any justification, but our aim here is to try to reduce its physical significance out of
theoretical reasons which invalidate Einstein's arguments in some probably realistic spacetime regions.
To be more precise, the geodesics of the Christoffel connection as the orbits of all particles
arise by the identification of the local inertial frame/freely falling frame
(ingeniously discovered by Einstein refining the notion of weak equivalence principle) with the local
Lorentz frame/local Minkowski frame (today the existence of such a frame probably does not need any
justification since the notion of a spacetime metric which generalizes the Minkowski metric is
considered fundamental, but it seems as well that it can consistently be introduced through what is
called Einstein equivalence principle concerning the non-gravitational laws of physics).

In this work we prove that in the presence of only massless particles (massive particles do not exist)
the above identification is meaningless and does not lead to any connection. In an alternative
construction of a connection, responsible for providing the orbits of all particles,
the existence of the above two frames can usually be adopted, but not their identification.
Our proof is based on the existence of projectively equivalent connections and the absence of
proper time for massless particles. The fact that the relevant group for massless particles is the
conformal group does not affect the validity of our result since the Lorentz group is still present.
We argue that this result is probably relevant in our world. It seems that the most direct and solid
regime for the applicability of the above result is the early universe. In the early universe,
the standard approach is to consider the contributions from loops and thermal corrections at
finite temperature. As a result, all particles are expected to be massless before the electroweak
symmetry breaking. Even for theories with phase transitions at higher energy scales, before these
scales all particles are still expected massless. If these massless particles have to decide about
their motion in a gravitational environment out of some kinematical principles (at least as long as
semiclassical notions such as orbits are still defined), they cannot be based on the notion of a
massive particle which arises later in time. Thus, the Christoffel connection in this regime is not
the result of physical arguments on particle kinematics, thus probably losing its overall reliability
(in this same spirit, even the axiom of continuation to the purely massless case of the minimal
length geodesics of the massive particles appeared later in time, is also insufficient).
On the contrary, if another connection could be defined out of physical arguments concerning the
kinematics of all particles, which arguments will equally well be meaningful everywhere in spacetime,
we consider that this new connection could be a more appropriate and realistic connection to be valid
everywhere in spacetime. Moreover, the existence of the notion of orbit (either classical or
semiclassical) seems still reasonable around the $\rm{TeV}$ scale since the full strong quantum
gravity regime is normally expected at much higher energies. Another case where the insufficiency
of the Christoffel connection may occur is in spacetime regions of very high energy and temperature
where only massless particles may exist, which regions are causally disconnected from the rest of
spacetime where massive particles are also present (deep interior of supermassive astrophysical
objects or even black holes are not impossible to be such regions). The above thoughts could also
be relevant in the presence of extra dimensions with some fundamental mass scale.

\section{The local inertial frames}
\label{inertial}

The local indistinguishability of gravitational and inertial forces, which is an essential step for
the derivation of Einstein's General Relativity (and therefore of other theories as well), is
described by the axiom that there exists a coordinate system $\chi^{\mu}$ around the event $p$
(local inertial frame or freely falling frame) and a parameter $\lambda$, such that the orbits
$\chi^{\mu}(\lambda)$ of freely falling test particles obey
\begin{equation}
\frac{d^{2} \chi^{\mu}}{d\lambda^{2}}(p)=0\,.
\label{jute}
\end{equation}
The orbit $\chi^{\mu}(\lambda)$ has locally the form of a straight line up to terms of
order $\mathcal{O}(\lambda^{3})$. Equation (\ref{jute}) is an infinitesimal version of the law
of inertia and implies the universality of free fall. It encapsulates some form of the weak
equivalence principle as refined by Einstein due to his epiphany by October 1907: ``I was sitting
in a chair in the patent office at Bern ...''. It turns out that $\lambda$ is an affine
parameter of a geodesic. In arbitrary coordinates $x^{\mu}$, equation (\ref{jute}) takes the form
\begin{equation}
\frac{d^{2}x^{\mu}}{d\lambda^{2}}+\omega^{\mu}_{\,\,\,\nu\kappa}
\frac{dx^{\nu}}{d\lambda}\frac{dx^{\kappa}}{d\lambda}=0
\label{gtde}
\end{equation}
at $p$, where
\begin{equation}
\omega^{\mu}_{\,\,\,\nu\kappa}=
\frac{\partial x^{\mu}}{\partial \chi^{\rho}}
\frac{\partial^{2} \chi^{\rho}}{\partial x^{\nu}\partial x^{\kappa}}\,.
\label{hywm}
\end{equation}
The quantity $\omega^{\mu}_{\,\,\,\nu\kappa}$ (symmetric in $\nu,\kappa$)
is easily seen that forms the components of a connection and is given by the functions
$\chi^{\mu}(x^{\nu})$. Formally, the analytical expression (\ref{hywm}) is defined not only
at $p$ but at other points as well. However, the correct is that the expression (\ref{hywm})
is meaningful and gives the values of the connection only at the point $p$. At another point
a corresponding expression will be valid for the corresponding $\chi^{\mu}$, and depending
on what assumptions one adopts, the form of the connection will arise at all points. Indeed,
if equation (\ref{hywm}) was valid in a neighborhood of $p$, applying this equation for the
specific system $x^{\mu}=\chi^{\mu}$, it would give a vanishing (flat) connection with vanishing
curvature, which certainly is not the situation that one wants to describe. For the same reason,
equation (\ref{jute}) cannot be true in a whole neighborhood around $p$ (something that sometimes is
misunderstood). However, equation (\ref{gtde}) for $x^{\mu}=\chi^{\mu}$ provides with the use of
(\ref{jute}) that at the point $p$ the components of the connection in the system $\chi^{\mu}$
vanish. So, basically the system $\chi^{\mu}$ defines the connection at $p$, such that at $p$
the components of the connection in its own coordinates $\chi^{\mu}$ vanish.

It is a crucial point that for massive particles the parameter $\lambda$ of (\ref{jute}) has for
consistency to be the proper time $\tau$ along the orbit of the particle (or more generally equal to
$\alpha\tau+\beta$, with $\alpha,\beta$ constants), defined by $d\tau^{2}=-g_{\mu\nu}dx^{\mu}dx^{\nu}$.
This is an assumption inside equation (\ref{jute}), and therefore, for massive particles the only
affine parameter is assumed to be the proper time. If this was not assumed, we will see that
inconsistency would appear even in our everyday world and the Christoffel connection would never arise.
For massless particles there is no proper time, so a parameter $\lambda$ is assumed to exist such
that equation (\ref{jute}) is still valid. In the massless case, the parameter $\lambda$ can be the
temporal coordinate $\chi^{0}$ since if $\lambda$ is different than $\chi^{0}$, one can define the
new coordinate system $\xi^{\mu}$ such that its spatial coordinates are $\xi^{i}=\chi^{i}$, while
$\xi^{0}=\lambda$, and then $\frac{d^{2}\xi^{\mu}}{d(\xi^{0})^{2}}(p)=0$. We will let a general
$\lambda$ satisfying equation (\ref{jute}) without some other specification,
except if stated otherwise explicitly. Another point to be noticed is the following: the arbitrary
system $x^{\mu}$ can equally well describe the motion of either massive or massless particles
(for example, the standard coordinates $t,r,\theta,\phi$ of a spherically symmetric configuration
describes all particles). Intuitively, this is so since the orbit of a massless particle is very
close to the orbit of a massive particle chasing the first one very closely. Since the connection
is a spacetime function, independent of the orbits, the inertial system $\chi^{\mu}$ of
(\ref{hywm}) describes all particles.

If, instead of $\lambda$, an arbitrary parameter $\sigma$ is used, then equation
(\ref{jute}) becomes
\begin{equation}
\frac{d^{2} \chi^{\mu}}{d\sigma^{2}}(p)=f(p)\frac{d\chi^{\mu}}{d\sigma}(p)\,,
\label{bwom}
\end{equation}
where
\begin{equation}
f=-\frac{d^{2}\sigma}{d\lambda^{2}}\Big(\frac{d\sigma}{d\lambda}\Big)^{\!\!-2}\,,
\label{jvhs}
\end{equation}
while equation (\ref{gtde}) becomes
\begin{equation}
\frac{d^{2}x^{\mu}}{d\sigma^{2}}+\omega^{\mu}_{\,\,\,\nu\kappa}\frac{dx^{\nu}}{d\sigma}
\frac{dx^{\kappa}}{d\sigma}=f\frac{dx^{\mu}}{d\sigma}
\label{niet}
\end{equation}
at $p$. The connection $\omega^{\mu}_{\,\,\,\nu\kappa}$ remains unchanged when we pass
from the affine parameter $\lambda$ to the non-affine $\sigma$. Equations (\ref{gtde}),
(\ref{niet}), as known, transform tensorially under coordinate transformations with $f$ unchanged.
As for the function $f$, it could in principle be different from geodesic to geodesic, depending on
the selected parametrization on each one.

If a coordinate system $\hat{x}^{\mu}$ satisfies
$\frac{\partial^{2}\hat{x}^{\mu}}{\partial \chi^{\nu}\partial\chi^{\kappa}}(p)=0$,
while the parameter $\hat{\lambda}$ obeys $\frac{d^{2}\lambda}{d\hat{\lambda}^{2}}(p)=0$,
then $\frac{d^{2}\hat{x}^{\mu}}{d\hat{\lambda}^{2}}(p)=0$. Thus, it seems that
$\chi^{\mu}(\lambda)$ may not be unique around $p$ with the property that ``eats up''
gravity, but other systems $\hat{x}^{\mu}(\hat{\lambda})$ could also exist with the same
property. Indeed, for example, the globally rotated systems
$\mathcal{X}^{\mu}=\gamma^{\mu}_{\,\,\,\nu}\chi^{\nu}$, where $\gamma^{\mu}_{\,\,\,\nu}$
a general invertible constant matrix, are also locally inertial with the same affine parameter
$\lambda$, and satisfy (\ref{jute}), (\ref{hywm}). In the following of this section,
we will show the existence of other non-trivial such systems and construct them more systematically
using two methods. In the first method, from an arbitrary system $x^{\mu}$, another local inertial
system $\tilde{x}^{\mu}$ will be constructed for any sort of particles being present, such that
$\tilde{x}^{\mu}(\lambda)$ satisfies (\ref{jute}) and also $\tilde{x}^{\mu}$ can replace
$\chi^{\mu}$ in (\ref{hywm}). In this sense, another system $y^{\mu}$ will define its own local
inertial frame $\tilde{y}^{\mu}$, different in general from $\tilde{x}^{\mu}$, such that again
$\tilde{y}^{\mu}(\lambda)$ satisfies (\ref{jute}), (\ref{hywm}), and so on. The system
$\tilde{y}^{\mu}(y^{\nu})$ can also be expressed in terms of $x^{\mu}$ as
$\tilde{y}^{\mu}(y^{\rho}(x^{\nu}))$. The inertial systems and the way they are constructed in the
first method do not lead to any inconsistency even if only massless particles are
present. They rather convince us in a more direct and manifest manner about the multiplicity of
inertial frames and they lead to the Christoffel connection. In the second method, from an arbitrary
but specific system $x^{\mu}$, in the case where only massless particles are present, we will
construct an infinity of local inertial frames $\bar{x}^{\mu}(x^{\nu})$ such that each one of them,
along with its appropriate parameter $\bar{\lambda}$, satisfies (\ref{jute}), i.e.
$\bar{x}^{\mu}(\bar{\lambda})$ satisfies (\ref{jute}), while it also satisfies a deformation of
(\ref{hywm}). The systems $\bar{x}^{\mu}$ are different in general from the systems
$\chi^{\mu},\mathcal{X}^{\mu},\tilde{x}^{\mu},\tilde{y}^{\mu},...$ because the affine parameter
$\bar{\lambda}$ is different than $\lambda$. If a massive particle is present, the systems
$\bar{x}^{\mu}$ reduce to $\tilde{x}^{\mu}$, since a $\bar{x}^{\mu}$ has to describe
both massive and massless particles. The existence of the systems $\bar{x}^{\mu}$ in the
purely massless case is the root for the proof of the claimed inconsistency.

For an arbitrary coordinate system $x^{\mu}$, we define the coordinate system
$\tilde{x}^{\mu}$ by the relation
\begin{equation}
x^{\mu}-x^{\mu}(p)=\tilde{x}^{\mu}-\tilde{x}^{\mu}(p)-\frac{1}{2}\omega^{\mu}_{\,\,\,\nu\kappa}(p)
(\tilde{x}^{\nu}-\tilde{x}^{\nu}(p))(\tilde{x}^{\kappa}-\tilde{x}^{\kappa}(p))\,.
\label{kneh}
\end{equation}
Maybe a few $\tilde{x}^{\mu}$ are defined, and not just one, due to the quadratic form of the
defining equation. As $\chi^{\mu}$, the system $\tilde{x}^{\mu}$, due to equation (\ref{kneh}),
is also valid for all particles. The components $\tilde{\omega}^{\mu}_{\,\,\,\nu\kappa}$ of the
connection in the coordinates $\tilde{x}^{\mu}$ are easily seen to have
$\tilde{\omega}^{\mu}_{\,\,\,\nu\kappa}(p)=0$. Therefore, since equations (\ref{gtde}),
(\ref{niet}) hold for any system, applying them for $x^{\mu}=\tilde{x}^{\mu}$ we get the corresponding
of (\ref{jute}), (\ref{bwom}), i.e.
\begin{eqnarray}
&&\frac{d^{2} \tilde{x}^{\mu}}{d\lambda^{2}}(p)=0\label{jybu}\\
&&\frac{d^{2} \tilde{x}^{\mu}}{d\sigma^{2}}(p)=f(p)\frac{d\tilde{x}^{\mu}}{d\sigma}(p)\,.
\label{bhby}
\end{eqnarray}
Moreover, a direct calculation shows that the expression (\ref{hywm}) is also valid with
$\tilde{x}^{\mu}$ playing the role of $\chi^{\mu}$, i.e. it is
\begin{equation}
\omega^{\mu}_{\,\,\,\nu\kappa}=
\frac{\partial x^{\mu}}{\partial \tilde{x}^{\rho}}
\frac{\partial^{2} \tilde{x}^{\rho}}{\partial x^{\nu}\partial x^{\kappa}}
\label{hbek}
\end{equation}
at $p$.{\footnote{Inversely, starting from a geodesic equation of the form (\ref{niet})
(with $f$ the same for all coordinate systems) for an arbitrary torsionless connection
$\omega^{\mu}_{\,\,\,\nu\kappa}$, equations (\ref{jybu})-(\ref{hbek}) arise at $p$ for the
system $\tilde{x}^{\mu}$ of (\ref{kneh}). Therefore, if a geodesic equation of some connection
provides the orbits of free particles, equation (\ref{jute}) is not anything fundamental, but it
is a simple geometric fact that is always true; in the hands of Einstein, however, equation
(\ref{jute}) obtained a primitive role which allowed him to determine the connection.}}
Obviously, due to (\ref{hbek}), the connection at $p$ defined by a
system $\tilde{x}^{\mu}$ coincides with $\omega^{\mu}_{\,\,\,\nu\kappa}(p)$.
These mean that for each coordinate system $x^{\mu}$, a corresponding local
inertial system $\tilde{x}^{\mu}$ can be defined with the same affine parameter $\lambda$.
It can be noticed that equation (\ref{kneh}) defining $\tilde{x}^{\mu}$ can be
written equivalently in exactly the same form if $x^{\mu}$ is replaced by
$X^{\mu}=\gamma^{\mu}_{\,\,\,\nu}x^{\nu}$, $\omega^{\mu}_{\,\,\,\nu\kappa}$ by the corresponding
components
$\Omega^{\mu}_{\,\,\,\nu\kappa}=\omega^{\rho}_{\,\,\,\sigma\tau}\gamma^{\mu}_{\,\,\,\rho}
(\gamma^{-1})^{\sigma}_{\,\,\,\nu}(\gamma^{-1})^{\tau}_{\,\,\,\kappa}$
of the same connection in the $X^{\mu}$ coordinates, and
$\tilde{x}^{\mu}$ by $\tilde{X}^{\mu}=\gamma^{\mu}_{\,\,\,\nu}\tilde{x}^{\nu}$.
The system $\tilde{X}^{\mu}$ satisfies (\ref{jybu}), (\ref{bhby}), while
$\Omega^{\mu}_{\,\,\,\nu\kappa}$, $X^{\mu}$, $\tilde{X}^{\mu}$ satisfy (\ref{hbek}).
Therefore, the (trivial) globally rotated frames of $\tilde{x}^{\mu}$ are locally inertial with
the same affine parameter $\lambda$ and arise from the class of the globally rotated systems of
$x^{\mu}$, still within the formula (\ref{kneh}).
Since equations (\ref{jybu})-(\ref{hbek}) are analogous to equations (\ref{jute}), (\ref{bwom}),
(\ref{hywm}), someone might think that the systems $\tilde{x}^{\mu}$ are not something new, but
they coincide with $\chi^{\mu}$ (or their rotations $\mathcal{X}^{\mu}$). However, this is not
true in general. Moreover, $\tilde{x}^{\mu}$ is not the rotation of $\tilde{y}^{\mu}$ in
general. Therefore, various $\tilde{x}^{\mu}$ exist, which are different from $\chi^{\mu}$ and
different from each other. To see this, let $x^{\mu}$ defines through (\ref{kneh}) the system
$\tilde{x}^{\mu}$, while another system $y^{\mu}$ defines $\tilde{y}^{\mu}$.
Let us suppose that $\tilde{x}^{\mu}$ coincides with $\gamma^{\mu}_{\,\,\,\nu}\chi^{\nu}$
and $\tilde{y}^{\mu}$ coincides with $\Gamma^{\mu}_{\,\,\,\nu}\chi^{\nu}$, where
$\gamma^{\mu}_{\,\,\,\nu},\Gamma^{\mu}_{\,\,\,\nu}$ are invertible constant matrices. Then,
$\tilde{y}^{\mu}=\Delta^{\mu}_{\,\,\,\nu}\tilde{x}^{\nu}$, where
$\Delta^{\mu}_{\,\,\,\nu}=\Gamma^{\mu}_{\,\,\,\kappa}(\gamma^{-1})^{\kappa}_{\,\,\,\nu}$.
This relation between $\tilde{x}^{\mu},\tilde{y}^{\mu}$ is shown not to be generic, even
if $\tilde{x}^{\mu},\tilde{y}^{\mu}$ are not related to $\chi^{\mu}$. Indeed, equation (\ref{kneh})
is a quadratic algebraic equation for $\tilde{x}^{\mu}$ and its solution for $\tilde{x}^{\mu}$ is
some algebraic expression containing square roots. Similarly, we solve the second equation
with $y^{\mu}$ for $\tilde{x}^{\mu}$; it will have another form with square roots with
some differences due to the different values of the connection in the $y^{\mu}$ system and the
possible existence of the free parameters $\Delta^{\mu}_{\,\,\,\nu}$ wandering around. Equating
the two expressions for $\tilde{x}^{\mu}$ we get that the system $y^{\mu}$ is related to the system
$x^{\mu}$ through a specific algebraic expression with square roots, containing also various
numerical values. Of course, such an expression does not exhaust an arbitrary coordinate system
$y^{\mu}$, so the above relation of $\tilde{x}^{\mu}$, $\tilde{y}^{\mu}$ only rarely occurs.
Therefore, formula (\ref{kneh}) defines the first method mentioned above for the determination
of various inertial frames.

There are two non-trivial things happening for the geodesic equation (\ref{niet}) with an arbitrary
$\omega^{\mu}_{\,\,\,\nu\kappa}$. First, equation (\ref{niet}) is quasi-form invariant under changes
of the parameter $\sigma$, something that does not happen in all ordinary differential equations,
although a change of the independent variable is always a permissible transformation. This precisely
means that the form of equation (\ref{niet}) does not change if $\sigma$ is replaced by
$\check{\sigma}$ and $f$ by $\check{f}$, where $\check{f}
=\big(f\frac{d\check{\sigma}}{d\sigma}\!-\!\frac{d^{2}\check{\sigma}}{d\sigma^{2}}\big)
\big(\frac{d\check{\sigma}}{d\sigma}\big)^{\!-2}$ (and to be more
precise $x^{\mu}$ by $x^{\mu}\circ \sigma$). Therefore, when we have a connection
$\omega^{\mu}_{\,\,\,\nu\kappa}$ at hand (in a coordinate system), the choice of the function
$f$ does not affect the geodesics (their point sets). So, while the new geodesic equation with
$\check{f}$ to be solved is different than the original one with $f$, the geodesics are the same.
The above expression for $\check{f}$ is in agreement with (\ref{jvhs}) setting $\check{f}=0$ and
$\check{\sigma}=\lambda$.

Second, if $\bar{q}^{\mu}$ is an arbitrary vector field, then
\begin{equation}
\omega_{_{\!\!\!\!\!\sim}\,\,\,\,\nu\kappa}^{\mu}=
\omega^{\mu}_{\,\,\,\nu\kappa}+\delta^{\mu}_{\nu}\bar{q}_{\kappa}+
\delta^{\mu}_{\kappa}\bar{q}_{\nu}
\label{ywmg}
\end{equation}
defines another torsionless connection. The value of $\omega_{_{\!\!\!\!\!\sim}\,\,\,\,\nu\kappa}^{\mu}$
at $p$ depends on the value of $\bar{q}^{\mu}$ at $p$ and no differentiation of $\bar{q}^{\mu}$
or $\omega^{\mu}_{\,\,\,\nu\kappa}$ will appear. However, the values of $\bar{q}^{\mu}$ along
the geodesic will be used in order to make changes in the parameter $\sigma$ or $\lambda$,
similarly to the fact that the various coordinate systems are also defined around $p$. The geodesic
equation (\ref{niet}) for $\omega^{\mu}_{\,\,\,\nu\kappa}$ is written as
\begin{equation}
\frac{d^{2}x^{\mu}}{d\sigma^{2}}+\omega_{_{\!\!\!\!\!\sim}\,\,\,\,\nu\kappa}^{\mu}
\frac{dx^{\nu}}{d\sigma}
\frac{dx^{\kappa}}{d\sigma}=\bar{f}\frac{dx^{\mu}}{d\sigma}
\label{jeot}
\end{equation}
at $p$, where
\begin{equation}
\bar{f}=f+2\bar{q}_{\rho}\frac{dx^{\rho}}{d\sigma}\,.
\label{nstm}
\end{equation}
Moreover, defining $\bar{\sigma}$ by the equation
\begin{equation}
\frac{d^{2}\bar{\sigma}}{d\sigma^{2}}+f\Big(\frac{d\bar{\sigma}}{d\sigma}\Big)^{2}-
\bar{f}\frac{d\bar{\sigma}}{d\sigma}=0\,,
\label{lbvo}
\end{equation}
equation (\ref{jeot}) becomes
\begin{equation}
\frac{d^{2}x^{\mu}}{d\bar{\sigma}^{2}}+\omega_{_{\!\!\!\!\!\sim}\,\,\,\,\nu\kappa}^{\mu}
\frac{dx^{\nu}}{d\bar{\sigma}}\frac{dx^{\kappa}}{d\bar{\sigma}}=f\frac{dx^{\mu}}{d\bar{\sigma}}
\label{kvie}
\end{equation}
at $p$. Alternatively, equation (\ref{kvie}) can arise by first performing a parameter change from
$\sigma$ to $\bar{\sigma}$ in (\ref{niet}) and then introducing
$\omega_{_{\!\!\!\!\!\sim}\,\,\,\,\nu\kappa}^{\mu}$ from (\ref{ywmg}). In the context of geometry,
the previous equations for $\omega_{_{\!\!\!\!\!\sim}\,\,\,\,\nu\kappa}^{\mu}$ are valid not only
at $p$ as here, but along the whole curves, and then, since (\ref{jeot}) is also the geodesic
equation of $\omega_{_{\!\!\!\!\!\sim}\,\,\,\,\nu\kappa}^{\mu}$, this means that the geodesics of
$\omega_{_{\!\!\!\!\!\sim}\,\,\,\,\nu\kappa}^{\mu}$, $\omega^{\mu}_{\,\,\,\nu\kappa}$ coincide
($\omega_{_{\!\!\!\!\!\sim}\,\,\,\,\nu\kappa}^{\mu}$ is called projectively equivalent to
$\omega^{\mu}_{\,\,\,\nu\kappa}$). So, two projectively equivalent connections have different $f$'s
with the same parametrization (i.e. $\sigma$), or alternatively, they can have the same $f$ with
different parametrizations (i.e. $\sigma,\bar{\sigma}$). In practice, equation (\ref{kvie}) for
suitable $\bar{q}^{\mu},\bar{\sigma}$ can facilitate the finding of the geodesics. Hence, the connection
$\omega_{_{\!\!\!\!\!\sim}\,\,\,\,\nu\kappa}^{\mu}$ defined by (\ref{ywmg}) is in some loose sense
a sort of ``gauge'' transformation for the geodesics. For our purposes here, the validity of
the previous equations at $p$ will be enough. The inverse is also true, which means that any two
connections which have all their geodesics the same are necessarily related by an equation of the
form (\ref{ywmg}) \cite{Eisen}, \cite{Ehl-Sch}. Indeed, if $v^{\mu}=\frac{dx^{\mu}}{d\sigma}$, the
geodesic equations of the projectively equivalent connections
$\omega^{\mu}_{\,\,\,\nu\kappa},\omega_{_{\!\!\!\!\!\sim}\,\,\,\,\nu\kappa}^{\mu}$ are written
as $\frac{dv^{\mu}}{d\sigma}+\omega^{\mu}_{\,\,\,\nu\kappa}v^{\nu}v^{\kappa}=hv^{\mu}$,
$\frac{dv^{\mu}}{d\sigma}+\omega_{_{\!\!\!\!\!\sim}\,\,\,\,\nu\kappa}^{\mu}v^{\nu}v^{\kappa}
=h_{_{\!\!\!\!\!\sim}}v^{\mu}$ for some appropriate functions $h,h_{_{\!\!\!\!\!\sim}}$.
Therefore $\bar{\Delta}_{\mu\nu\kappa}v^{\nu}v^{\kappa}=Hv_{\mu}$, where
$\bar{\Delta}_{\mu\nu\kappa}=
\omega_{\mu\nu\kappa}-\omega_{_{_{\!\!\!\!\!\sim}}\mu\nu\kappa}=\bar{\Delta}_{\mu\kappa\nu}$,
$H=h-h_{_{\!\!\!\!\!\sim}}$. The above algebraic condition for $v^{\mu}$ implies that there are
vector fields $A^{\mu}$, $B^{\mu}$, such that
$\bar{\Delta}_{\mu\nu\kappa}=g_{\mu\nu}A_{\kappa}+g_{\mu\kappa}B_{\nu}$. The symmetry property of
$\bar{\Delta}_{\mu\nu\kappa}$ implies $A_{\mu}=B_{\mu}$, and thus equation (\ref{ywmg}) arises.

We now focus to the case where only massless particles are present. We will find other local
inertial frames, for which the equivalence principle holds, in a different way. Namely,
using a single coordinate system $x^{\mu}$, we will define various systems $\bar{x}^{\mu}$
(other than the global rotations of the corresponding system $\tilde{x}^{\mu}$) such that still
gravity is locally eliminated. If in particular we attach with such a $\bar{x}^{\mu}$ the
appropriate affine parameter $\bar{\lambda}$, we will have the form (\ref{jute}) for
$\bar{x}^{\mu}(\bar{\lambda})$. Indeed, we have defined in (\ref{lbvo}) the parameter
$\bar{\sigma}$, such that the geodesic equation for
$\omega_{_{\!\!\!\!\!\sim}\,\,\,\,\nu\kappa}^{\mu}$ at $p$ takes the specific form (\ref{kvie}).
Define now the parameter $\bar{\lambda}$ by the equation
\begin{equation}
f=\frac{d^{2}\bar{\lambda}}{d\bar{\sigma}^{2}}\Big(\frac{d\bar{\lambda}}{d\bar{\sigma}}\Big)^{\!-1}
\label{lvek}
\end{equation}
and we get
\begin{equation}
\frac{d^{2}x^{\mu}}{d\bar{\lambda}^{2}}+\omega_{_{\!\!\!\!\!\sim}\,\,\,\,\nu\kappa}^{\mu}
\frac{dx^{\nu}}{d\bar{\lambda}}\frac{dx^{\kappa}}{d\bar{\lambda}}=0
\label{nwlo}
\end{equation}
at $p$. The coordinate system $\bar{x}^{\mu}$ defined by
\begin{equation}
x^{\mu}-x^{\mu}(p)=
\bar{x}^{\mu}-\bar{x}^{\mu}(p)-\frac{1}{2}\omega_{_{\!\!\!\!\!\sim}\,\,\,\,\nu\kappa}^{\mu}(p)
(\bar{x}^{\nu}-\bar{x}^{\nu}(p))(\bar{x}^{\kappa}-\bar{x}^{\kappa}(p))
\label{nhmj}
\end{equation}
has vanishing components at $p$ of the $\omega_{_{\!\!\!\!\!\sim}}$ connection in the
$\bar{x}^{\mu}$ system, i.e. $\bar{\omega}_{_{\!\!\!\!\!\sim}\,\,\,\,\nu\kappa}^{\mu}(p)=0$,
quite similarly to what happens with equation (\ref{kneh}). Due to (\ref{nhmj}), the system
$\bar{x}^{\mu}$ is valid for all particles, massive and massless, similarly to what happens with
the systems $\tilde{x}^{\mu}$. Since equation (\ref{nwlo}) holds for any system, applying it
for $x^{\mu}=\bar{x}^{\mu}$, we get the corresponding of (\ref{jute}), i.e.
\begin{equation}
\frac{d^{2} \bar{x}^{\mu}}{d\bar{\lambda}^{2}}(p)=0\,.
\label{lmko}
\end{equation}
Moreover, it arises
\begin{equation}
\omega_{_{\!\!\!\!\!\sim}\,\,\,\,\nu\kappa}^{\mu}=
\frac{\partial x^{\mu}}{\partial \bar{x}^{\rho}}
\frac{\partial^{2} \bar{x}^{\rho}}{\partial x^{\nu}\partial x^{\kappa}}
\label{sier}
\end{equation}
at $p$. These mean that various local inertial coordinate systems $\bar{x}^{\mu}$, each one with
an appropriate affine parameter $\bar{\lambda}$, arise from the projectively equivalent connections
$\omega_{_{\!\!\!\!\!\sim}\,\,\,\,\nu\kappa}^{\mu}$ (as parametrized by $\bar{q}^{\mu}$). Due to
(\ref{sier}), the connection at $p$ defined by a system $\bar{x}^{\mu}$ coincides with
$\omega_{_{\!\!\!\!\!\sim}\,\,\,\,\nu\kappa}^{\mu}(p)$.

Since the role played by the parameter $\bar{\lambda}$ is crucial, it is enlightening to simplify
the previous analysis in order to make the situation even more clear. We consider in (\ref{jeot})
the affine parameter $\lambda$ instead of the general parameter $\sigma$ and get
\begin{equation}
\frac{d^{2}x^{\mu}}{d\lambda^{2}}+\omega_{_{\!\!\!\!\!\sim}\,\,\,\,\nu\kappa}^{\mu}
\frac{dx^{\nu}}{d\lambda}
\frac{dx^{\kappa}}{d\lambda}=2\bar{q}_{\rho}\frac{dx^{\rho}}{d\lambda}
\frac{dx^{\mu}}{d\lambda}
\label{hynr}
\end{equation}
at $p$. Equation (\ref{hynr}) also arises directly from (\ref{gtde}) introducing the quantity
$\omega_{_{\!\!\!\!\!\sim}\,\,\,\,\nu\kappa}^{\mu}$. Then, we define directly the parameter
$\bar{\lambda}$ from the equation
\begin{equation}
\frac{d^{2}\bar{\lambda}}{d\lambda^{2}}
=2\bar{q}_{\rho}\frac{dx^{\rho}}{d\lambda}\frac{d\bar{\lambda}}{d\lambda}\,.
\label{hndm}
\end{equation}
Equation (\ref{hndm}) also arises from the previous equations, since for $f=0$ in (\ref{lvek}),
a solution can be $\bar{\lambda}=\bar{\sigma}$, and equation (\ref{lbvo}) provides (\ref{hndm}).
Finally, changing in (\ref{hynr}) from $\lambda$ to $\bar{\lambda}$ according to (\ref{hndm}),
we get (\ref{nwlo}). We can summarize by saying that adding in (\ref{gtde}) the $\bar{q}_{\mu}$
terms in order to get the projectively equivalent connection
$\omega_{_{\!\!\!\!\!\sim}\,\,\,\,\nu\kappa}^{\mu}$ as in (\ref{hynr}), a non-affine geodesic
arises which can be changed into the affine one (\ref{nwlo}) through (\ref{hndm}). Equations
(\ref{lmko}), (\ref{sier}) follow exactly as before. The parameter $\bar{\lambda}$ is different
from $\lambda$ (or $\alpha\lambda+\beta$ with $\alpha,\beta$ constants), otherwise equation
(\ref{hndm}) would be inconsistent. So, the process is that a $\bar{q}^{\mu}$ defines a
$\omega_{_{\!\!\!\!\!\sim}\,\,\,\,\nu\kappa}^{\mu}$ through (\ref{ywmg}) and a $\bar{\lambda}$
through (\ref{hndm}), then a $\bar{x}^{\mu}$ is defined through (\ref{nhmj}) and satisfies
equations (\ref{lmko}), (\ref{sier}). The above description shows that instead of changing the
coordinate systems to get other local inertial frames, we can stay at a single coordinate system
and just flip to projectively equivalent connections in order to get new
local inertial frames. Therefore, formula (\ref{nhmj}) defines the second method mentioned above for
the determination of inertial frames in the purely massless case.

As in the case with the systems $\tilde{x}^{\mu}$, also here, equation (\ref{nhmj}) defining
$\bar{x}^{\mu}$ can be written equivalently in exactly the same form if $x^{\mu}$ is replaced by
$X^{\mu}=\gamma^{\mu}_{\,\,\,\nu}x^{\nu}$, $\omega_{_{\!\!\!\!\!\sim}\,\,\,\,\nu\kappa}^{\mu}$
by the corresponding components
$\Omega_{_{\!\!\!\!\!\sim}\,\,\,\,\nu\kappa}^{\mu}=\omega_{_{\!\!\!\!\!\sim}\,\,\,\,\sigma\tau}^{\rho}
\gamma^{\mu}_{\,\,\,\rho}(\gamma^{-1})^{\sigma}_{\,\,\,\nu}(\gamma^{-1})^{\tau}_{\,\,\,\kappa}$
of the $\omega_{_{\!\!\!\!\!\sim}}$ connection in the $X^{\mu}$ coordinates, and
$\bar{x}^{\mu}$ by $\bar{X}^{\mu}=\gamma^{\mu}_{\,\,\,\nu}\bar{x}^{\nu}$.
The system $\bar{X}^{\mu}$ satisfies (\ref{lmko}), while
$\Omega_{_{\!\!\!\!\!\sim}\,\,\,\,\nu\kappa}^{\mu}$, $X^{\mu}$, $\bar{X}^{\mu}$ satisfy (\ref{sier}).
Therefore, the (trivial) globally rotated frames of $\bar{x}^{\mu}$ are locally inertial with
affine parameter $\bar{\lambda}$ and arise from the class of the globally rotated systems of
$x^{\mu}$, still within the formula (\ref{nhmj}). The systems $\bar{x}^{\mu}$ are different in
general (i) from the systems $\tilde{x}^{\mu}$ or $\tilde{X}^{\mu}$, and (ii) from
$\mathcal{X}^{\mu}$ and from each other. As for (i), this is seen by combining
equations (\ref{kneh}), (\ref{nhmj}) and $\tilde{x}^{\mu}=\gamma^{\mu}_{\,\,\,\nu}\bar{x}^{\nu}$,
from where it arises
$[\gamma^{\mu}_{\,\,\,\rho}\omega^{\rho}_{\,\,\,\nu\kappa}(p)
-\Omega_{_{\!\!\!\!\!\sim}\,\,\,\,\nu\kappa}^{\mu}(p)]
[\tilde{x}^{\nu}-\tilde{x}^{\nu}(p)][\tilde{x}^{\kappa}-\tilde{x}^{\kappa}(p)]=
2(\gamma^{\mu}_{\,\,\,\nu}-\delta^{\mu}_{\,\,\,\nu})[\tilde{x}^{\nu}-\tilde{x}^{\nu}(p)]$. There is no
$\gamma^{\mu}_{\,\,\,\nu}$ such that the last equation is valid, since the left hand side is
quadratic in $\tilde{x}^{\mu}-\tilde{x}^{\mu}(p)$ while the right hand side is linear,
and also $\tilde{x}^{\mu}$ depends on $x^{\mu}$. The same fact can also be seen as follows. The
solution of equation (\ref{hndm}) will in general have both $\frac{d^{2}\bar{\lambda}}{d\lambda^{2}}(p)$,
$\frac{d\bar{\lambda}}{d\lambda}(p)$ non-vanishing, since the opposite is a very special initial
condition for the differential equation (\ref{hndm}), which means a very special choice for the
solution $\bar{\lambda}(\lambda)$. If $\tilde{x}^{\mu}=\gamma^{\mu}_{\,\,\,\nu}\bar{x}^{\nu}$,
then equations (\ref{lmko}), (\ref{jybu}) would give $\frac{d^{2}\bar{\lambda}}{d\lambda^{2}}(p)=0$,
which is incompatible in general. As for (ii), let a $\bar{q}^{\mu}$
defines $\omega_{_{\!\!\!\!\!\sim}\,\,\,\,\nu\kappa}^{\mu}$ and $\bar{x}^{\mu}$, while another
$\bar{\bar{q}}^{\mu}$ defines $\omega_{_{\!\!\!\!\!\sim_{\!\!\!\!\sim}}\,\,\,\,\nu\kappa}^{\mu}$
and $\bar{\bar{x}}^{\mu}$. If $\bar{x}^{\mu}=\gamma^{\mu}_{\,\,\,\nu}\chi^{\nu}$ and
$\bar{\bar{x}}^{\mu}=\Gamma^{\mu}_{\,\,\,\nu}\chi^{\nu}$, then
$\bar{\bar{x}}^{\mu}=\Delta^{\mu}_{\,\,\,\nu}\bar{x}^{\nu}$, where
$\Delta^{\mu}_{\,\,\,\nu}=\Gamma^{\mu}_{\,\,\,\kappa}(\gamma^{-1})^{\kappa}_{\,\,\,\nu}$.
This relation between $\bar{x}^{\mu},\bar{\bar{x}}^{\mu}$ is not generic, even if
$\bar{x}^{\mu},\bar{\bar{x}}^{\mu}$ are not related to $\chi^{\mu}$. Indeed, from equation
(\ref{nhmj}) applied twice for $\bar{x}^{\mu},\bar{\bar{x}}^{\mu}$, we get
$[\omega_{_{\!\!\!\!\!\sim_{\!\!\!\!\sim}}\,\,\,\,\rho\sigma}^{\mu}(p)\Delta^{\rho}_{\,\,\,\nu}
\Delta^{\sigma}_{\,\,\,\kappa}-\omega_{_{\!\!\!\!\!\sim}\,\,\,\,\nu\kappa}^{\mu}(p)]
[\bar{x}^{\nu}-\bar{x}^{\nu}(p)][\bar{x}^{\kappa}-\bar{x}^{\kappa}(p)]=
2(\Delta^{\mu}_{\,\,\,\nu}-\delta^{\mu}_{\,\,\,\nu})[\bar{x}^{\nu}-\bar{x}^{\nu}(p)]$.
As before, there is no $\Delta^{\mu}_{\,\,\,\nu}$ such that the last equation is valid, since the
left hand side is quadratic in $\bar{x}^{\mu}-\bar{x}^{\mu}(p)$ while the right hand side is linear,
and also $\bar{x}^{\mu}$ depends on $x^{\mu}$. The same conclusion is also derived from
equation (\ref{hndm}) applied twice for $\bar{\lambda},\bar{\bar{\lambda}}$, and we get
$\frac{d^{2}\bar{\bar{\lambda}}}{d\bar{\lambda}^{2}}\big(\frac{d\bar{\lambda}}
{d\lambda}\big)^{2}=2(\bar{\bar{q}}_{\rho}-\bar{q}_{\rho})\frac{dx^{\rho}}
{d\lambda}\frac{d\bar{\bar{\lambda}}}{d\lambda}$. If
$\bar{\bar{x}}^{\mu}=\Delta^{\mu}_{\,\,\,\nu}\bar{x}^{\nu}$, then equation (\ref{lmko}) applied
twice for $\bar{x}^{\mu},\bar{\bar{x}}^{\mu}$ provides
$\frac{d^{2}\bar{\bar{\lambda}}}{d\bar{\lambda}^{2}}(p)=0$, and therefore
$\bar{\bar{q}}^{\mu}(p)=\bar{q}^{\mu}(p)$, which is inconsistent in general.

When massive particles are not present, the root of the claimed inconsistency lies on the
existence of the local inertial frames $\bar{x}^{\mu}$ which give rise to the projectively
equivalent connections $\omega_{_{\!\!\!\!\!\sim}\,\,\,\,\nu\kappa}^{\mu}$ according to (\ref{sier}).
If, however, massive particles are present, the existence of proper time for the massive particles
changes completely the situation and the systems $\bar{x}^{\mu}$ reduce to $\tilde{x}^{\mu}$.
Equations (\ref{nwlo}), (\ref{lmko}) for the massive particles should have
$\bar{\lambda}=\lambda=\tau$ (or $\alpha\tau+\beta$) since the weak equivalence principle in
the form (\ref{jute}) has embedded the assumption of proper time. Although the affine parameters
between massive and massless particles are different, the system $\bar{x}^{\mu}$ should be the same
for both. Equation (\ref{hndm}) for a massive particles provides $\bar{q}^{\mu}=0$, thus
$\omega_{_{\!\!\!\!\!\sim}\,\,\,\,\nu\kappa}^{\mu}=\omega^{\mu}_{\,\,\,\nu\kappa}$
and $\bar{x}^{\mu}=\tilde{x}^{\mu}$. Therefore, if massive particles are present, the Christoffel
connection arises normally and no inconsistency occurs. The same conclusion arises, although in a
little more complicated way, if we think the presence of massive particles in terms of the general
non-affine parameter $\sigma$ instead of the proper time. Indeed, equation (\ref{lvek}) defines
the parameter $\bar{\sigma}(\tau)$. Since $\bar{\lambda}=\lambda=\tau$,
equations (\ref{lvek}), (\ref{jvhs}) coincide, and thus $\bar{\sigma}=\sigma$ (up to a possible
additive constant). Hence, from equation (\ref{lbvo}) it arises $\bar{f}=f$ and (\ref{nstm})
gives $\bar{q}^{\mu}=0$. In other words, a general
$\omega_{_{\!\!\!\!\!\sim}\,\,\,\,\nu\kappa}^{\mu}$, different than $\omega^{\mu}_{\,\,\,\nu\kappa}$,
satisfies for massive particles equation (\ref{hynr}) with a non-vanishing right hand side, in
contrast to equation (\ref{gtde}), which means that
$\bar{x}^{\mu}$ satisfies (\ref{lmko}) and not the axiom of the weak equivalence principle for
massive particles in the form (\ref{jute}). Finally, let us finish with a hypothetical comment:
If $\bar{x}^{\mu}$ for timelike particles was different than $\bar{x}^{\mu}$ for null particles
(which is not the case), then the construction of the systems $\bar{x}^{\mu}$ would invalidate
the derivation of the Christoffel connection even in the presence of massive particles.

\section{The local Lorentz frames}

Given a spacetime metric $g_{\mu\nu}$, for a given point $p$ there is a surrounding coordinate system
$x'^{\mu}$, such that the corresponding components have $g'_{\mu\nu}(p)=\eta_{\mu\nu}$. This is an
issue of linear algebra since the matrix $g_{\mu\nu}(p)$ can always be diagonalized and $x'^{\mu}$
can be constructed such that the vectors $\partial/\partial x'^{\mu}$ at $p$ coincide with the
orthonormal vectors of diagonalization. Therefore, an infinite number of such systems $x'^{\mu}$
exist. Indeed, any system $x''^{\mu}$, such that $x''^{\mu}-x''^{\mu}(p)$ equals
$x'^{\mu}-x'^{\mu}(p)$ plus quadratic or higher powers of $x'^{\mu}-x'^{\mu}(p)$, has
$g_{\mu\nu}''(p)=\eta_{\mu\nu}$. While the weak equivalence principle intents to embody gravity,
a metric with spacetime signature intends to embrace special relativity. Although the local
Minkowski structure is always present (basically through the tangent space) and a coordinate
system exists such that the metric locally coincides with the Minkowski metric, obviously there
are coordinate systems around $p$ with $g_{\mu\nu}(p)\neq \eta_{\mu\nu}$. In an arbitrary coordinate
system $x^{\mu}$ it holds
\begin{equation}
g_{\mu\nu}=\frac{\partial x'^{\rho}}{\partial x^{\mu}}
\frac{\partial x'^{\sigma}}{\partial x^{\nu}}\eta_{\rho\sigma}
\label{kner}
\end{equation}
at $p$. This equation cannot be
true in a whole neighborhood around $p$, since then, the space would be Minkowski in a whole neighborhood,
which certainly is not the situation that one wants to describe. Today, a metric $g_{\mu\nu}$ is
considered as a fundamental object, but it was not always obvious that such a $g_{\mu\nu}$ should exist
and generalize the Minkowski metric in the presence of gravity. It seems that what is called Einstein
equivalence principle implies the existence of a $g_{\mu\nu}$ which is locally Minkowski
\footnote{ The Hughes-Drever experiment rules out the existence of more than one second-rank tensor
field both coupling directly to matter, however, vector and tensor fields which couple only to
gravity or to matter's self gravitational energy are not ruled out \cite{Will-Nordtvedt},
\cite{Will-Hawking-Israel}.}. This principle extends the weak equivalence principle and requires
the independency from the direction and spacetime position for any freely falling frame (where
inhomogeneities of the external fields are ignorable) of any local non-gravitational test experiment.
Thus, the metric arises as a direct result of the existence of the freely falling frames together
with their direction insensitivity (local Lorentz invariance of the non-gravitational laws of
physics) and position independency (see \cite{Will}, p. 23 for more details). However, the selection
of the Christoffel connection is not automatic and has to be justified.

Another coordinate system $x'^{^{\!\!\!\!\!\circ}\,\,\mu}$, more specific than $x'^{\mu}$, can be
defined through the relation
\begin{equation}
x'^{\mu}-x'^{\mu}(p)=x'^{^{\!\!\!\!\!\circ}\,\,\mu}-
x'^{^{\!\!\!\!\!\circ}\,\,\mu}(p)-\frac{1}{2}
\Gamma'^{\mu}_{\,\,\,\,\,\nu\kappa}(p)(x'^{^{\!\!\!\!\!\circ}\,\,\nu}-
x'^{^{\!\!\!\!\!\circ}\,\,\nu}(p))
(x'^{^{\!\!\!\!\!\circ}\,\,\kappa}-x'^{^{\!\!\!\!\!\circ}\,\,\kappa}(p))\,,
\label{brue}
\end{equation}
where $\Gamma'^{\mu}_{\,\,\,\,\,\nu\kappa}$ are the components of the Christoffel connection in the
system $x'^{\mu}$. Then, it is easily seen that $g'^{^{\!\!\!\!\!\circ}}_{\mu\nu}(p)=\eta_{\mu\nu}$
and $g'^{^{\!\!\!\!\!\circ}}_{\mu\nu}(x'^{^{\!\!\!\!\!\circ}\,\,\kappa})$ has vanishing first
derivatives at $p$, i.e.
$\frac{\partial g'^{^{\!\!\!\!\!\circ}}_{\mu\nu}}{\partial x'^{^{\!\!\!\!\!\circ}\,\,\kappa}}(p)=0$,
$\Gamma'^{^{^{\!\!\!\!\!\!\circ}}\,\,\,\mu}_{\,\,\,\,\,\,\nu\kappa}(p)=0$ (the small circle above the
corresponding symbols reminds us notationally of these zero values). Also, equation (\ref{kner}) holds
at $p$ with $x'^{^{\!\!\!\!\!\circ}\,\,\mu}$ taking the place of $x'^{\mu}$. These are simple geometrical
facts which are generally true and are sometimes useful for performing proofs in tensor analysis.
A system $x'^{^{\!\!\!\!\!\circ}\,\,\mu}$ is called local Lorentz frame (or local Minkowski
frame) around $p$. Therefore, gravitation, which is manifested through $g_{\mu\nu}$ (and possibly
other fields), is a second-order effect. Certainly the existence of coordinate systems such as
$x'^{\mu}$, or even more $x'^{^{\!\!\!\!\!\circ}\,\,\mu}$, expresses the fact that a gravity theory
agrees locally with special relativity to a good approximation. In an arbitrary coordinate
system $x^{\mu}$ it is
\begin{equation}
g_{\mu\nu}=\frac{\partial x'^{^{\!\!\!\!\!\circ}\,\,\rho}}{\partial x^{\mu}}
\frac{\partial x'^{^{\!\!\!\!\!\circ}\,\,\sigma}}{\partial x^{\nu}}
g'^{^{\!\!\!\!\!\circ}}_{\rho\sigma}\,.
\label{nhra}
\end{equation}
Differentiation of (\ref{nhra}) gives
\begin{equation}
\frac{\partial g_{\mu\nu}}{\partial x^{\kappa}}=\eta_{\rho\sigma}
\frac{\partial x'^{^{\!\!\!\!\!\circ}\,\,\rho}}{\partial x^{\mu}}
\frac{\partial^{2}x'^{^{\!\!\!\!\!\circ}\,\,\sigma}}
{\partial x^{\nu}\partial x^{\kappa}}+\eta_{\rho\sigma}
\frac{\partial x'^{^{\!\!\!\!\!\circ}\,\,\rho}}{\partial x^{\nu}}
\frac{\partial^{2}x'^{^{\!\!\!\!\!\circ}\,\,\sigma}}{\partial x^{\mu}\partial x^{\kappa}}
\label{jueb}
\end{equation}
at $p$.

If we consider the (global) Lorentz transformation $x'\!\!\!\!\!\!-^{\,\mu}=
\Lambda^{\mu}_{\,\,\,\nu}x'^{\nu}$ of $x'^{\mu}$, then the corresponding metric
$g'\!\!\!\!\!\!-_{\mu\nu}$ also has the property $g'\!\!\!\!\!\!\!-_{\mu\nu}(p)=\eta_{\mu\nu}$.
Accordingly, the systems $x'^{^{\!\!\!\!\!\circ}\,\mu}\!\!\!\!\!\!\!\!-\,\,\,
=\Lambda^{\mu}_{\,\,\,\nu}x'^{^{\!\!\!\!\!\circ}\,\,\nu}$ satisfy
$\,\,g'^{^{\!\!\!\!\!\circ}}_{\mu\nu}\!\!\!\!\!\!\!\!\!\!\!-\,\,\,\,(p)=\eta_{\mu\nu}$,
$\frac{\partial g'^{^{\!\!\!\!\!\circ}}_{\mu\nu}\!\!\!\!\!\!\!\!\!\!-\,\,}
{\partial x'^{^{^{\!\!\!\!\!\circ}}\,\kappa}\!\!\!\!\!\!\!\!\!-\,\,}\,\,(p)=0$, so they define
other local Lorentz frames. Equation (\ref{brue}) defining $x'^{^{\!\!\!\!\!\circ}\,\,\mu}$
can be written equivalently in exactly the same form if $x'^{\mu}$ is replaced by
$x'\!\!\!\!\!\!-^{\,\mu}$, $\Gamma'^{\mu}_{\,\,\,\,\,\nu\kappa}$ by the corresponding
components $\Gamma'^{\rho}_{\,\,\,\,\,\sigma\tau} \Lambda^{\mu}_{\,\,\,\rho}
(\Lambda^{-1})^{\sigma}_{\,\,\,\nu}(\Lambda^{-1})^{\tau}_{\,\,\,\kappa}$ of the Christoffel
connection in the $x'\!\!\!\!\!\!-^{\,\mu}$ coordinates, and $x'^{^{\!\!\!\!\!\circ}\,\,\mu}$
by $x'^{^{\!\!\!\!\!\circ}\,\mu}\!\!\!\!\!\!\!\!-\,\,\,$. Therefore, the globally rotated
systems of $x'^{^{\!\!\!\!\!\circ}\,\,\mu}$ arise from the class of the globally rotated
systems of $x'^{\mu}$, still within the formula (\ref{brue}). Equations (\ref{nhra}), (\ref{jueb})
remain the same with $x'^{^{\!\!\!\!\!\circ}\,\mu}\!\!\!\!\!\!\!\!-\,\,$ substituting
$x'^{^{\!\!\!\!\!\circ}\,\,\mu}$.

\section{The proof of the inconsistency}

The notion of connection (which determines the free fall of particles) and the notion
of metric (which embodies the local Minkowski structure) are in general unrelated.
It was Einstein's ingenuity to relate them in order to determine the connection
$\omega^{\mu}_{\,\,\,\nu\kappa}$. So, the Christoffel connection arises by the
identification of a freely falling frame $\chi^{\mu}$ or $\tilde{x}^{\mu}$ with a local Lorentz frame
$x'^{^{\!\!\!\!\!\circ}\,\,\mu}$, i.e.
$\chi^{\mu}=x'^{^{\!\!\!\!\!\circ}\,\,\mu}$ or $\tilde{x}^{\mu}=x'^{^{\!\!\!\!\!\circ}\,\,\mu}$.
Substituting the second derivatives in (\ref{jueb}) from (\ref{hywm}) or (\ref{hbek}), and making
use of (\ref{kner}) for $x'^{^{\!\!\!\!\!\circ}\,\,\mu}$, it arises
\begin{equation}
\frac{\partial g_{\mu\nu}}{\partial x^{\kappa}}=g_{\mu\sigma}\omega^{\sigma}_{\,\,\,
\nu\kappa}+g_{\nu\sigma}\omega^{\sigma}_{\,\,\,\mu\kappa}
\label{gbek}
\end{equation}
at $p$. This is the known equation of vanishing non-metricity and for a torsionless connection
it gives, after some algebraic manipulation, the Christoffel connection,
\begin{equation}
\omega^{\mu}_{\,\,\,\nu\kappa}=\Gamma^{\mu}_{\,\,\,\nu\kappa}
\label{ngas}
\end{equation}
at $p$. The connection was found to be the Christoffel connection at the point $p$, and since the same
can be repeated for any point, finally the Christoffel connection arises globally.

If massive particles are present (possibly together with massless), no systems $\bar{x}^{\mu}$
exist other than $\tilde{x}^{\mu}$, as explained in Sec. \ref{inertial}, due to the existence of
proper time assumed in the weak equivalence principle for massive particles. Thus, nothing more can
be said beyond equation (\ref{ngas}) which is the final word for the connection. For a massive
particle with $\lambda=\tau$ we have the obvious equation
$\frac{d}{d\lambda}\big(g_{\mu\nu}\frac{dx^{\mu}}{d\lambda}\frac{dx^{\nu}}{d\lambda}\big)=0$.
The left hand side of this equation, after making use of (\ref{gtde}) for the connection
(\ref{ngas}), turns out to be identically zero. This is a consistency check for the assumption
$\lambda=\tau$ in (\ref{jute}).

On the other hand, if only massless particles are present, the freely falling frames $\chi^{\mu}$,
$\tilde{x}^{\mu}$ still exist, but now the extra local inertial frames $\bar{x}^{\mu}$ are also
present. Each one of all these systems should be identified with some local Lorentz frame
$x'^{^{\!\!\!\!\!\circ}\,\,\mu}$. For $\chi^{\mu}=x'^{^{\!\!\!\!\!\circ}\,\,\mu}$ or
$\tilde{x}^{\mu}=x'^{^{\!\!\!\!\!\circ}\,\,\mu}$, equations (\ref{gbek}), (\ref{ngas}) at $p$
arise in exactly the same way as before. However, the unavoidable identification of $\bar{x}^{\mu}$
with some $x'^{^{\!\!\!\!\!\circ}\,\,\mu}$, i.e. $\bar{x}^{\mu}=x'^{^{\!\!\!\!\!\circ}\,\,\mu}$,
gives some extra conditions. Indeed, substituting the second derivatives in (\ref{jueb}) from
(\ref{sier}), and making use of (\ref{kner}) for $x'^{^{\!\!\!\!\!\circ}\,\,\mu}$, it arises
\begin{equation}
\frac{\partial g_{\mu\nu}}{\partial x^{\kappa}}=g_{\mu\sigma}
\omega_{_{\!\!\!\!\!\sim}\,\,\,\,\nu\kappa}^{\sigma}
+g_{\nu\sigma}\omega_{_{\!\!\!\!\!\sim}\,\,\,\,\mu\kappa}^{\sigma}
\label{nhgt}
\end{equation}at $p$. Again, algebraic manipulation of equation (\ref{nhgt}), as before, provides
that $\omega_{_{\!\!\!\!\!\sim}\,\,\,\,\nu\kappa}^{\mu}$ is the Christoffel connection, i.e
\begin{equation}
\omega_{_{\!\!\!\!\!\sim}\,\,\,\,\nu\kappa}^{\mu}=\Gamma^{\mu}_{\,\,\,\nu\kappa}
\label{kbbb}
\end{equation}
at $p$. Combining equations (\ref{ngas}), (\ref{kbbb}) gives
$\delta^{\mu}_{\nu}\bar{q}_{\kappa}+\delta^{\mu}_{\kappa}\bar{q}_{\nu}=0$ at $p$.
Multiplication of this equation with $\bar{q}_{\mu}$ gives $\bar{q}_{\nu}\bar{q}_{\kappa}=0$
at $p$. This last equation obviously gives $\bar{q}^{\mu}=0$ at $p$. To get this result even more
manifestly, since $\bar{q}^{\mu}$ is arbitrary, there are various vector fields $\bar{q}^{\mu}$
with non-vanishing magnitude at $p$, i.e. $\bar{q}^{\mu}\bar{q}_{\mu}\neq 0$ at $p$.
Considering equation $\bar{q}_{\nu}\bar{q}_{\kappa}=0$ for such a $\bar{q}^{\mu}$
and multiplying with $\bar{q}^{\kappa}$, we get $\bar{q}_{\nu}=0$. Equation $\bar{q}_{\nu}=0$ at
$p$ is certainly inconsistent since $\bar{q}^{\mu}$ is arbitrary. Therefore, we have proved that
the identification of a local inertial frame with a local Lorentz frame in the purely massless
case is meaningless and does not provide any connection (Christoffel or other). The reason for
this inconsistency is the absence of proper time and the existence of a distinct affine parameter
$\bar{\lambda}$, a different one for each local inertial frame $\bar{x}^{\mu}$, which satisfies
equation (\ref{lmko}). The local inertial systems $\bar{x}^{\mu}(\bar{\lambda})$ are equally good
as the local inertial systems $\chi^{\mu}(\lambda)$ or $\tilde{x}^{\mu}(\lambda)$ which satisfy
equations (\ref{jute}), (\ref{jybu}). The infinite number of the local inertial systems
$\bar{x}^{\mu}$ arise from the infinite number of the projectively equivalent connections
$\omega_{_{\!\!\!\!\!\sim}\,\,\,\,\nu\kappa}^{\mu}$, which are parametrized by the vector field
$\bar{q}^{\mu}$. The inconsistency is due to that any such connection turns out to coincide with
the Christoffel connection. If for massive particles it was not assumed that the affine parameter
is the proper time, then it becomes obvious that the same inconsistency would occur along
equations (\ref{nhgt}), (\ref{kbbb}) even in the presence of massive particles.

\section{Conclusions}
\label{Conclusions}

The assumption of the existence of a local inertial coordinate system around a spacetime point,
for which the weak equivalence principle is valid, automatically implies the existence of
infinitely many other such systems. The identification of any such system with a local
Lorentz coordinate system implies that the Christoffel connection controls the kinematics
of all particles. However, if only massless particles are present, the situation changes
drastically and we show that the identification of a freely falling frame with a local Lorentz
frame is meaningless and no connection arises. The presence of the conformal structure for massless
particles, instead of the Lorentz one, does not influence our result.

The reason for this inconsistency is that the absence of proper time in the purely massless
case allows for the existence of even more local inertial coordinate systems
$\bar{x}^{\mu}$. Technically, these last systems arise from the existence of projectively
equivalent connections $\omega_{_{\!\!\!\!\!\sim}\,\,\,\,\nu\kappa}^{\mu}$ through
the expression (\ref{nhmj}) and are parametrized by a vector field $\bar{q}^{\mu}$ along the
formula (\ref{ywmg}). The appropriate affine parameter $\bar{\lambda}$ of these systems for
the null particles is defined from (\ref{hndm}) and equation (\ref{lmko}) of free motion
is derived, as well as the corresponding geodesic equation (\ref{nwlo}).
The inconsistency arises from equation (\ref{sier}), when it is attempted to identify
$\bar{x}^{\mu}$ with a local Lorentz frame, since all the connections
$\omega_{_{\!\!\!\!\!\sim}\,\,\,\,\nu\kappa}^{\mu}$ turn out to coincide with the Christoffel
one.

Christoffel connection is theoretically significant in physics not because it is introduced as
an axiom, or because it is economical and solely constructed in terms of the metric, but
because it arises as a result of basic ideas concerning particle kinematics. If there are spacetime
regions where only massless particles exist and no massive particles are present to move
in the interior of the null cone, the Christoffel connection cannot arise out of the
above particle kinematics, and therefore, this connection probably loses its overall
significance as an exact connection of spacetime. A connection arising from different ideas of
particle kinematics, which are meaningful in the whole of spacetime, is probably more satisfactory
conceptually.

In the early universe, before spontaneous symmetry breaking at electroweak scale, all particles
are expected to be massless. In this region, which is not expected to be the full quantum
gravity regime, notions such as orbits are still meaningful classically or semiclassically.
The massless particles in this spacetime region have to decide about their motion in a gravitational
environment, but they cannot be based on the notion of a massive particle which appears
later in time. Therefore, the kinematical emergence of Christoffel connection in this region
breaks down and can set in doubt its overall reliability. Other spacetime regions with
only massless particles could be imagined, although probably with not no strong arguments, such as
the very interior of astrophysical objects or black holes, or relevant objects in the presence of
extra dimensions.

Let us finish with an interesting remark. Suppose we have a gravity theory where a connection (or
some of its components) is part of the dynamical fields. There are three options. In the first, one
insists, due to Einstein's kinematical arguments, that the motion of particles is still governed by
the Christoffel connection and that the other connection only indirectly influences the orbits through
its interaction with the metric. However, this reasoning is insufficient whenever the previously
shown inconsistency of the derivation of Christoffel connection is valid. In the second option,
the orbit is governed by the geodesic equation of the connection which caries torsional or
non-metricity degrees of freedom. In general, this equation will not preserve in time the nullity
constraint of a massless particle, except if this demand is appropriately taken into account,
possibly in the construction of the connection. In the third option, the motion of the particles
is provided by some action. This action can either refer to extended fields which give the
equation of motion for the particle through the geometric optics limit, or it can be a
delta-like action providing the orbit directly. Beyond that there are many correction terms
that can be added in such actions, similarly to the second case, the respect to the nullity condition
only occasionally is expected to occur, if not built-in inside the action.

\begin{acknowledgments}
I would like to thank I. Dalianis, A. Kehagias, E. Kiritsis, M. Maggiore, T. Sotiriou, T. Tomaras,
R. Troncoso and V. Zarikas for useful discussions.

\end{acknowledgments}


\end{document}